\pdfoutput=1 
\documentclass[aps,prl,twocolumn,superscriptaddress,amsfont,graphicx,nofootinbib,preprintnumbers]{revtex4}
\usepackage{color,graphicx,epsfig}
\usepackage{ifpdf}
\usepackage{amsmath}
\usepackage{bm}
\usepackage{color}
\usepackage[english]{babel}
\usepackage{graphicx}
\usepackage{amsfonts}
\usepackage{amssymb}
\usepackage{braket}
\usepackage{hyperref}
\usepackage{enumerate}

\definecolor{nicered}{rgb}{0.7,0.1,0.1}
\definecolor{nicegreen}{rgb}{0.1,0.5,0.1}
\hypersetup{colorlinks,citecolor= nicegreen,linkcolor= nicered}

\newcommand{\ep}{\epsilon}
\newcommand{\slashed}{\slash \hspace{-0.19cm}}

\newcommand{\be}{\begin{equation}}
\newcommand{\ee}{\end{equation}}
\newcommand{\bea}{\begin{eqnarray}}
\newcommand{\eea}{\end{eqnarray}}

\definecolor{Red}{rgb}{1.,0.,0.}

\DeclareMathOperator{\Li}{Li}

\newcommand{\oT}{\overline{T}}
\newcommand{\oF}{\overline{\mathcal{F}}}

\def\OMIT#1{}

\definecolor{darkred}{rgb}{0.9,0,0}

\definecolor{darkgreen}{rgb}{0,0,0.9}

\definecolor{darkblue}{rgb}{0,0,0.9}

\begin{document}

\def\OX{Rudolf Peierls Centre for Theoretical Physics, University of Oxford, Clarendon Laboratory, Parks
Road, Oxford OX1 3PU}
\def\MSU{Department of Physics and Astronomy, Michigan State University, East Lansing, Michigan 48824, USA}
\def\WDM{Wadham College, University of Oxford, Parks Road, Oxford OX1 3PN, UK}

\preprint{MSUHEP-20-018, OUTP-20-12P}

\title{Di-photon amplitudes in three-loop Quantum Chromodynamics}

\author{Fabrizio Caola}            
\email[Electronic address:]{fabrizio.caola@physics.ox.ac.uk}
\affiliation{\OX} 
\affiliation{\WDM}

\author{Andreas von Manteuffel}            
\email[Electronic address:]{vmante@msu.edu}
\affiliation{\MSU}

\author{Lorenzo Tancredi}            
\email[Electronic address:]{lorenzo.tancredi@physics.ox.ac.uk}
\affiliation{\OX}

\begin{abstract}
We consider the three-loop scattering amplitudes for the production of a pair of photons in quark-antiquark
annihilation in Quantum Chromodynamics (QCD).
We use suitably defined projectors to 
efficiently calculate all helicity amplitudes.
We obtain relatively compact analytic results,
that we write in terms of harmonic polylogarithms
or, alternatively, multiple polylogarithms of up to depth three.
This is the first calculation 
of a three-loop four-point scattering amplitude in 
full QCD.
\end{abstract}

\maketitle
Multiloop scattering amplitudes in Quantum Chromodynamics (QCD) 
are a crucial ingredient for the precision physics program carried out at 
particle colliders, such as the  Large Hadron Collider (LHC) at CERN. 
When combined with the corresponding real radiation, 
they make it possible to compute
high-precision predictions for a multitude of key processes.
In the recent past, such calculations played a fundamental role for Higgs 
characterisation~\cite{deFlorian:2016spz}, for the extraction of fundamental parameters
of the SM~\cite{Aaboud:2017svj,Ball:2017nwa} and for the study of electroweak bosons~\cite{Sirunyan:2020pub,Sirunyan:2020jtq,ATLAS:2020xqa}
and the top quark~\cite{ATLAS:2014aaa}, to mention a few examples. These investigations
allow for 
a deep scrutiny of the Standard Model (SM), which is essential for establishing
its validity and for revealing possible tensions pointing towards new physics.  

Together with their importance for particle collider phenomenology, 
scattering amplitudes in QCD also provide us with an important laboratory to investigate
formal properties of the perturbative expansion of realistic Quantum Field Theories (QFTs).
With the increase of the perturbative order, one faces the computation
of increasingly complicated multiloop Feynman diagrams which can depend on
multiple scales according to the number of particles involved in the collision.
This translates into increasingly involved analytic properties of the special
functions required to express the corresponding scattering amplitudes in closed
analytic form. 
The calculation of multi-loop scattering amplitudes have contributed to reveal
intricate and fascinating mathematical structures, 
whose further investigation has the potential to
improve our understanding of perturbative QFT.

For these reasons, in the last two decades a considerable effort has been devoted to push SM calculations further.
As a result, an impressive set of techniques have been developed,
which have made it possible to compute most $2 \to 2$ scattering amplitudes
up to two loops in massless QCD for many process relevant at hadron colliders~\cite{Garland:2002ak,Anastasiou:2002zn,Glover:2003cm,Glover:2004si,Bern:2001df,Bern:2002tk,Bern:2003ck,Gehrmann:2011ab,Gehrmann:2011aa,Gehrmann:2013vga,Caola:2014iua,Caola:2015ila,Gehrmann:2015ora,vonManteuffel:2015msa} and also to include massive effects both
approximately and exactly~\cite{Bonciani:2010mn,Baernreuther:2013caa,Borowka:2016ehy,Melnikov:2017pgf,Kudashkin:2017skd,Baglio:2018lrj,Heller:2019gkq,Frellesvig:2019byn,Bronnum-Hansen:2020mzk} . More recently, also the first results for
two-loop amplitudes for $2 \to 3$ scattering processes in QCD have been obtained~\cite{Abreu:2017hqn,Abreu:2018zmy,Abreu:2018jgq,Chicherin:2018yne,Badger:2019djh,Chicherin:2019xeg,Abreu:2020cwb,DeLaurentis:2020qle}, 
opening the way to the first
$2 \to 3$ Next-to-Next-to-Leading-Order (NNLO) studies at the LHC~\cite{Chawdhry:2019bji,Kallweit:2020gcp}.

In parallel to the effort to overcome the barrier of two-loop scattering amplitudes for $2 \to 3$
processes, similar work is required to push the calculation of $2 \to 2$ QCD processes 
up to three loops. Indeed, while the first $2 \to 2$ three-loop
results have appeared in $\mathcal N=4$ Super Yang-Mills~\cite{Henn:2016jdu} and in $\mathcal N=8$ Super Gravity~\cite{Henn:2019rgj},
 three loop calculations in realistic QFTs have only been performed
 for simple $2 \to 1$ processes~\cite{Moch:2005tm,Baikov:2009bg,Gehrmann:2010ue}. Only recently some of the ingredients for obtaining three-loop
 $2\to2$  amplitudes in massless QCD have started to appear~\cite{Ahmed:2019qtg,Henn:2020lye}. 

In this letter, we document the calculation of the three-loop QCD corrections to the scattering amplitude for di-photon production in quark-antiquark annihilations.  
This is the first three-loop amplitude for a $2\to2$ scattering process in 
full-colour QCD. The
$q\bar q\to \gamma\gamma $ process is arguably a very natural place to start
the investigation of three-loop corrections to 
$2 \to 2$ scattering in a realistic QFT.
Indeed, these amplitudes involve only massless external and virtual particles and,
moreover,
have a much simpler colour structure than 
the corresponding amplitudes for the production
of  coloured partons.
The relevant one-loop and two-loop amplitudes have been
known for a long time~\cite{Bern:2001df,Glover:2003cm}, followed by a multitude of 
phenomenological studies up to NNLO in QCD~\cite{Bern:2002pv,Catani:2011qz,Gehrmann:2020oec}. 
However, we would like to stress that despite the aforementioned simplifications, this process still
contains all of the analytic complexity of a generic massless $2\to 2$ scattering process. Hence,
we expect that the results obtained here can be extended to compute all 
three-loop scattering amplitudes for the production of two massless partons in hadronic collisions.

We now describe our calculation. We consider the  process
\begin{equation}
q (p_1) + \bar{q}(p_2)  + \gamma(p_3) + \gamma(p_4) \to 0\,, \label{eq:qqaa}
\end{equation}
and obtain its physical equivalent $q\bar q\to\gamma\gamma$ via crossing
symmetry $p_{3,4}\to - p_{3,4}$. 
We parametrise the kinematics of   Eq.~\eqref{eq:qqaa} by defining the usual Mandelstam invariants
\begin{gather}
s = (p_1+ p_2)^2\,,~ t = (p_1+p_3)^2\,,~u = (p_2+p_3)^2,\notag\\
p_1^2=p_2^2=p_3^2=p_4^2=0,\quad s + t + u = 0\,.
\end{gather}
 We also introduce the dimensionless variable $x = - t/s$, such that 
 in the physical scattering region, one has
\begin{equation}
s>0\,, \;\; t < 0\,,\;\; u < 0\,\; \Rightarrow \; 0<x<1. \label{eq:reg}
\end{equation}

We use the helicity of the incoming quark $\lambda_q$ and of the incoming photons
$\lambda_{3,4}$ to label the scattering amplitude of the process. We denote the scattering amplitude between
well-defined helicity states by $\mathcal A_{\lambda_q\lambda_3\lambda_4}$. 
Using parity, charge-conjugation and symmetry relations, it is easy to see that there are only two independent
helicity configurations~\cite{Bern:2001df,Glover:2003cm}. In what follows, 
for clarity we will compute
the over-complete set of four helicity configurations 
$\{\lambda_q \lambda_3 \lambda_4\} = \{L--\},\{L-+\},\{L+-\},\{L++\}$,
which allow us to obtain the remaining ones for right-handed quarks 
by a simple charge-conjugation transformation, as it will be discussed below.

In order to compute the helicity amplitudes, we regulate both infrared and ultraviolet divergences using dimensional regularisation, i.e. we work in $d=4-2\ep$ dimensions.
By choosing a gauge such that
\begin{align}
\epsilon_i \cdot p_i = 0\,,\;\; i=3,4\;\;\mbox{and} \;\; \epsilon_3 \cdot p_2 = \epsilon_4 \cdot p_1 = 0\,, \label{eq:gauge}
\end{align}
it is easy to see that, at any order in QCD perturbation theory, 
Lorentz covariance dictates that 
the amplitude can be parametrised as
\begin{align}
\mathcal{A}(s,t) = \sum_{i=1}^5 \mathcal{F}_i(s,t)\, \bar{u}(p_2) \, \Gamma_i^{\mu \nu} \,u(p_1) \epsilon_{3,\mu} \epsilon_{4,\nu}
\label{eq:dec1}
\end{align}
where $u$ and  $\bar u$ are the spinors for the incoming quark and antiquark, respectively, and the five Lorentz tensors $\Gamma_i^{\mu \nu}$ are defined as
\begin{align}
\Gamma_1^{\mu \nu} &= \gamma^\mu p_2^{\nu}\,, \;\; \Gamma_2^{\mu \nu} = \gamma^\nu p_1^{\mu}\,, \nonumber \\
\Gamma_3^{\mu \nu} &=  \slashed{p}_3\,p_1^\mu p_2^\nu \,,\;\; 
\Gamma_4^{\mu \nu} =  \slashed{p}_3\,g^{\mu \nu}\,, \nonumber \\
& \quad \Gamma_5^{\mu \nu} =  \gamma^\mu \slashed{p}_3\, \gamma^\nu.
\end{align}
The functions $\mathcal{F}_i(s,t)$ are scalar form factors which only depend on the Mandelstam 
invariants and on the number of dimensions $d$.
Since the colour structure is straight-forward, we keep colour indices implicit here.
For ease of typing, we define the five independent  structures
\begin{equation}
T_i =  \bar{u}(p_2) \, \Gamma_i^{\mu \nu} \,u(p_1) \epsilon_{3,\mu} \epsilon_{4,\nu}\,,
\end{equation}
and from now on, with a slight abuse of language, we  refer to the $T_i$ as the independent \emph{tensors} for the 
problem at hand.

At first sight, it may seem puzzling that we find five independent tensor structure when we have only four 
 helicity amplitudes (not considering charge conjugation). This mismatch is however easy to explain: the decomposition
Eq.~\eqref{eq:dec1} is valid for arbitrary dimension $d$. For four-dimensional external states, 
the five tensor structures $T_i$ are no 
longer independent, and four of them are enough to span the whole space~\cite{Peraro:2019cjj,peraro2}. Since eventually we are only
interested in the $d\to 4$ limit, it is 
convenient to reorganise the tensors $T_i$ and choose for $T_5$
 a linear combination which is identically zero when four-dimensional external states are considered.
This can be achieved by choosing 
\begin{align}
&\oT_i = T_i\,, \;\; i=1,...,4 \,, \nonumber \\
&\oT_5 = T_5 - \frac{u}{s} T_1 + \frac{u}{s} T_2 - \frac{2}{s}T_3 + T_4\,.
\label{eq:barT5}
\end{align}
We then write the scattering amplitude as
\begin{align}
\mathcal{A}(s,t) = \sum_{i=1}^5 \oF_i(s,t)\, \oT_i\,, \label{eq:AT}
\end{align}
where the form factors $\oF_i$ are suitable linear combinations of the original $\mathcal{F}_i$,
whose explicit form will be irrelevant in the following. All the non-trivial  information for
the process of Eq.~\eqref{eq:qqaa} is encoded in the form factors $\oF_i$. 
We stress once more that, while 
all five form factors are in general required for the  result 
in Conventional Dimensional Regularisation (CDR),
only the first four are enough to compute the helicity amplitudes in tHV, 
see the supplement material for more details.

Eq.~\eqref{eq:AT} can be inverted to select individual form factors. 
This is done by introducing suitable projectors
operators defined such that\footnote{The individual form factors are 
\emph {not} gauge-invariant, see Eq.~\eqref{eq:gauge}, and 
the reference vector $q_{3(4)} = p_{2(1)}$ should be used in the polarisation sums.}
\begin{equation}
\sum_{pol} \mathcal P_i \mathcal A(s,t) = \oF_i(s,t)\,,
\end{equation}
where we use $d$ dimensional polarization sums.
Since the five $\oT_i$ form a basis for our space, we can write the projectors as
\begin{align}
\mathcal{P}_i = \sum_{k=1}^5 c_{ik} \oT_k^\dagger \label{eq:projT}\,.
\end{align}
A straightforward calculation, reported in the supplement material, shows that
the matrix $c_{ik}$ is block-diagonal
\begin{equation}
c_{ik} = \left[
\begin{array}{cc}
C_{4\times4} & 0  \\
0 &  c_{55}
\end{array}
\right]\,. \label{eq:matrixprojqqbaa}
\end{equation}
As a consequence, the fifth tensor $\oT_5$ decouples from the other four.
This, combined with the fact that it always evaluates to zero for four-dimensional external states,
shows that the helicity amplitudes receive contributions only from the first four form factors in Eq.~\eqref{eq:AT}. In other words, \emph{there is a one-to-one correspondence between helicity amplitudes
and form factors in the 't\ Hooft-Veltman scheme (tHV)}~\cite{tHooft:1972tcz}. 
This statement is quite general and 
it is always possible to construct a basis where the decoupling
of tensors leading to vanishing results in $d=4$ is manifest. 
Details of how to achieve this in general 
will be presented elsewhere~\cite{peraro2}.

In order to compute the form factors, we generate the
relevant tree-level, one-, two- and three-loop Feynman diagrams with
\texttt{QGRAF}~\cite{Nogueira:1991ex} and apply on each of them the the 
projectors defined in 
Eqs~(\ref{eq:projT}) and (\ref{eq:matrixprojqqbaa}). 
We perform all required colour and Dirac algebra algebra  
with \texttt{FORM}~\cite{Vermaseren:2000nd}. 
There are $2$ diagrams at
tree level, $10$ at one loop, $143$ at two loops and $2922$ at three loops.
At a given number of loops, each Feynman diagram can be identified  with a graph,
which is specified by the permutation of the external legs and by the number and topology 
of the internal lines.
In particular, at three loops, where the number of Feynman diagrams becomes rather large,  
it is convenient to sum together those diagrams that are associated to 
the same graph and manipulate them 
together to avoid as much as possible to perform the same operation multiple times.
Performing these steps allows us to express the three-loop form factors in terms of a 
large number of scalar integrals
of the form\footnote{Note that $n_i \in \mathbb{Z}$, i.e. the integrals in 
Eq.~\eqref{eq:genI}
can have irreducible numerator structures.}
\begin{align}
\mathcal{I}^{\rm fam}_{n_1,...,n_{15}} =  e^{3 \epsilon \gamma_E} \,\int \prod_{i=1}^3 \frac{d^d k_i}{i \pi^{d/2}} \frac{1}{D_1^{n_1} ... D_{15}^{n_{15}}}\,,\label{eq:genI}
\end{align}
with $\gamma_E \approx 0.5772$ the Euler constant.  
The propagators $D_j$ can be drawn from three different families of 
integrals ${\rm fam} = {\rm \{PL,NPL1,NPL2 \}}$ and their crossings. The definition of the three families
is provided in electronic files attached to this letter.
A key features of integrals of the form Eq.~\eqref{eq:genI} is that they are not all
independent. Instead, through a by-now standard
use of integration-by-parts identities, most of them can be expressed in terms of 
a much smaller set of so-called master integrals~\cite{Tkachov:1981wb,Chetyrkin:1981qh}.
This can be done in principle algorithmically for any process~\cite{Laporta:2001dd} for integrals of the form
Eq.~\eqref{eq:genI}. 
Despite being well-understood in principle~\cite{Laporta:2001dd}, this reduction step is computationally highly non-trivial for the problem at hand, as it requires the symbolic 
solution of large systems of linear equations. We have achieved it through
a combined use of the public code Reduze 2~\cite{vonManteuffel:2012np,Studerus:2009ye},
and an in-house implementation, Finred, employing finite field sampling~\cite{vonManteuffel:2014ixa,vonManteuffel:2016xki,Peraro:2016wsq,Peraro:2019svx}
and syzygy techniques~\cite{Gluza:2010ws,Schabinger:2011dz,Ita:2015tya,Larsen:2015ped,Boehm:2017wjc,Agarwal:2019rag}.
In this way, the three loop form factors $\oF_j^{(3)}(s,t)$ can be expressed as linear combinations of a total
of $486$ ($132$) independent master integrals, including (excluding) integrals related by crossings of the external legs.

A basis of master integrals, excluding integrals obtained from them by crossings of the external legs, 
has been computed for the first time in Ref.~\cite{Henn:2020lye} using the method of
differential equations~\cite{Bern:1993kr,Kotikov:1990kg,Remiddi:1997ny,Gehrmann:1999as,Primo:2016ebd} augmented by the choice of a canonical basis~\cite{Kotikov:2010gf,Henn:2013pwa}. There it was shown that
all these integrals can be expressed in terms of a particularly simple and well understood class of functions, harmonic 
polylogarithms (HPLs)~\cite{Goncharov:1998kja,Remiddi:1999ew,Vollinga:2004sn,Duhr:2011zq,Duhr:2012fh}. 
The present calculation requires results through to transcendental weight six.\footnote{Note that
a simple counting argument shows that there can be at 
most $146$ functions that can contribute at weight six. 
This puts an upper bound on the number of independent master integrals 
that can contribute to the finite part of the three-loop amplitude.}
We have re-derived 
the differential equations fulfilled by the master integrals
including all their non-trivial
crossings required for the calculation of the actual scattering amplitudes.
We have verified that the results of Ref.~\cite{Henn:2020lye} 
do fulfil the differential equations. As a check, we 
also have recomputed various boundary constants for the integrals of Ref.~\cite{Henn:2020lye} independently. 
For manipulating the harmonic polylogarithms, we used in-house routines and
\texttt{PolyLogTools}~\cite{Duhr:2019tlz}.

Extra care has to be put in computing the boundary conditions and in analytically continuing the integrals 
to the physical scattering region in Eq.~\eqref{eq:reg}.
Indeed, it is well known that scattering amplitudes
are multivalued complex functions of the external kinematics 
and that, in order to obtain physical results, one must 
define the integrals on the correct Riemann sheet 
whenever one or more of the Mandelstam invariants
are positive. 
The standard approach consists, whenever possible, in computing the relevant scattering amplitude for 
non-physical values of the kinematics where all Mandelstam variables are
negative and a real result is expected. If this is possible, it is then often simple to
analytically continue the amplitudes to physical kinematics using the Feynman prescription 
$s_i \to s_i + i 0$
for each Mandelstam invariant $s_i$ which crosses a branch cut.
Unfortunately, this approach fails for $2 \to 2$ massless scattering where the momentum
conservation relation $s+t+u=0$ prevents us from finding such a non-physical region.
As a consequence, the scattering amplitudes must be computed directly in a region where at least one 
Mandelstam invariant is positive.
In the present case, extra care has to be taken since the boundary conditions provided in Ref.~\cite{Henn:2020lye} 
are computed on the unphysical Riemann sheet 
 reached by giving a small and \emph{negative} imaginary part to the Mandelstam invariant $s \to s - i 0$. Of course, it is expected that 
 these results should be related to the ones 
 for $s \to s + i 0$ by complex conjugation.
We have continued the results of Ref.~\cite{Henn:2020lye} back to the physical Riemann
sheet and checked that this holds. 
We then used them to obtain analytic expressions for all required crossings (see~\cite{Anastasiou:2000mf} for details).
We used these integrals to compute the three loop  form factors $\oF^{(3)}_j(s,t)$. 

The results computed with the procedure discussed so far contain both ultraviolet and infrared singularities. 
Up to three loops, they can be written as follows
\be
\oF_i = \delta_{kl} (4 \pi \alpha)\,e_q^2\, \sum_{k=0}^{3} \left(\frac{\alpha_{s,b}}{2\pi}\right)^k\oF_{i}^{(k,b)},
\ee
where $e = \sqrt{4 \pi \alpha}$ is the electric charge, $e_q$ is the charge of the incoming quark in units of $e$, $\alpha_{s,b}$ is the bare strong coupling and $\delta_{kl}$ carries the colour indices of the two incoming quarks. We remove ultraviolet singularities by expressing our
result in terms of the $\overline{ \rm MS}$ renormalised coupling $\alpha_s(\mu)$: 
\be
\oF_i = \delta_{kl} (4 \pi \alpha)\,e_q^2\, \sum_{k=0}^{3} \left(\frac{\alpha_{s}(\mu)}{2\pi}\right)^k\oF_{i}^{(k)}.
\label{eq:renF}
\ee
The relation between renormalised and bare coupling is given by
\be
S_{\epsilon}\alpha_{s,b} = \mu^{2\ep} \alpha_s(\mu) Z[{\alpha_s(\mu)}],
\ee
with
$S_{\epsilon} = (4\pi)^{-\epsilon}e^{-\gamma_E \epsilon}$
and
\be
Z[\alpha] = 1-\frac{\beta_0}{\ep}\left(\frac{\alpha_s}{2\pi}\right) + 
\left(\frac{\beta_0^2}{\epsilon^2}-\frac{\beta_1}{2\epsilon}\right)
\left(\frac{\alpha_s}{2\pi}\right)^2+ \mathcal O(\alpha_s^3)\,.
\ee
The explicit form of the $\beta$-function coefficient $\beta_{0,1}$ 
is reported in the supplemental material. For definiteness, we will
present our results for
$\mu^2=s$. It is straightforward to obtain
results at any other scale using 
renormalisation-group arguments. 

The renormalised form factors $\oF_i^{(k)}$ of Eq.~\eqref{eq:renF} still contain infrared singularities. 
Their form 
is universal and can be expressed in terms of the lower-order scattering amplitudes as follows~\cite{Catani:1998bh,Becher:2009qa}
\be
\begin{split}
&
\oF_i^{(1)} = \mathcal I_1 \oF_i^{(0)} + \oF_i^{(1,{\rm fin})},
\\
&
\oF_i^{(2)} = \mathcal I_2\oF_i^{(0)} + \mathcal I_1 \oF_i^{(1)} + \oF_i^{(2,{\rm fin})},
\\
&
\oF_i^{(3)} = \mathcal I_3 \oF_i^{(0)} + \mathcal I_2 \oF_i^{(1)} + 
\mathcal I_1 \oF_i^{(2)} + \oF_i^{(3,{\rm fin})}.
\end{split}
\label{eq:catani}
\ee
In these equations, the $\mathcal I_j$ are universal factors that only depend on the center of mass energy of the coloured
partons $s$,  on $\ep$ and
on the QCD Casimirs $C_F = 4/3$, $C_A=3$, as well as on the number of light fermions $n_f$. We spell them out explicitly in the supplemental material. For our discussion, it is only important to note that
the function $\mathcal I_i$ contains infrared poles up to order $2^i$, i.e. $\mathcal I_1 \sim 1/\ep^2$,
$\mathcal I_2\sim 1/\ep^4$, $\mathcal I_3 \sim 1/\ep^6$. The finite remainders $\oF_i^{(k,{\rm fin)}}$ in
Eq.~\eqref{eq:catani} are
finite in the $\ep\to 0$ limit, and contain all the non-trivial physics information for the
process Eq.~\eqref{eq:qqaa}. 

As we have already mentioned, it is straightforward to obtain the helicity amplitudes from our form factors
by evaluating the tensor structures $\oT_i$ for well-defined helicity states, see Eq.~\eqref{eq:barT5}.
We stress once more that for any helicity configuration one has $\oT_{5,\lambda_q\lambda_3\lambda_4}=0$. 
We write for left-handed spinors $\bar{u}_L(p_2) = \langle 2 |$ and $u_L(p_1) = | 1 ]$
and for the the photon $j$ of momentum $p_j$ 
$$\epsilon^\mu_{j,-}(q_j) = \frac{\langle q_j | \gamma^\mu | j ] }{ \sqrt{2} \langle q_j j \rangle}\,, \quad
\epsilon^\mu_{j,+}(q_j) = \frac{\langle j | \gamma^\mu | q_j ] }{ \sqrt{2} [ j q_j  ]}\,.$$
With these definitions we find 
\begin{equation}
\begin{split}
    &\mathcal A_{L--} = \frac{2[34]^2}{\langle 1 3 \rangle [23]} 
    \alpha(x) \,, \quad 
    \mathcal A_{L-+} = \frac{2\langle 24 \rangle [13]}{\langle 2 3 \rangle [24]} 
     \beta(x)\,, \\
    &\mathcal A_{L+-} = \frac{2\langle 23 \rangle [41]}{\langle 2 4 \rangle [32]} 
    \gamma(x) \,, \quad
    \mathcal A_{L++} = \frac{2\langle 34\rangle^2}{\langle 3 1 \rangle [23]} 
    \delta(x) \,.
    \label{eq:helamp}
    \end{split}
\end{equation}
with
\begin{equation}
\begin{split}
&\alpha(x) =  \frac{t}{2}\,\left( \oF_2 - \frac{t}{2} \oF_3 +  \oF_4 \right) ,
\\ 
&\beta(x) = \frac{t}{2}\, \left( \frac{s}{2} \oF_3  +  \oF_4 \right) ,
\\
&\gamma(x) =  \frac{s\,t}{2u}\left( \oF_2 - \oF_1  - \frac{t}{2} \oF_3 - \frac{t}{s}\oF_4 \right)\,,
\\
&\delta(x) =\frac{t}{2}\left( \oF_1 + \frac{t}{2} \oF_3 - \oF_4 \right)\,.
\label{eq:alphabeta}
\end{split}
\ee
Note that the remaining 
amplitudes for right-handed quarks can be obtained by 
a charge-conjugation transformation as follows
\begin{equation}
    \mathcal{A}_{R \lambda_3 \lambda_4} = \mathcal{A}_{L \lambda^*_3 \lambda^*_4}\left( \langle ij \rangle \leftrightarrow [ji] \right)\,,
\end{equation}
where $\lambda_i^*$ indicates the opposite helicity of $\lambda_i$.
Symmetry under exchange of the two photons 
requires
\be
\label{eq:bosesymm}
\gamma(x)=\beta(1-x),~ \delta(x)=-\alpha(x),~\alpha(1-x) = -\alpha(x),
\ee
and at tree-level we find $\alpha=\delta=0$ and $\beta=\gamma=1$.

\begin{figure*}[th]
    \centering
    \includegraphics[width=.48\linewidth]{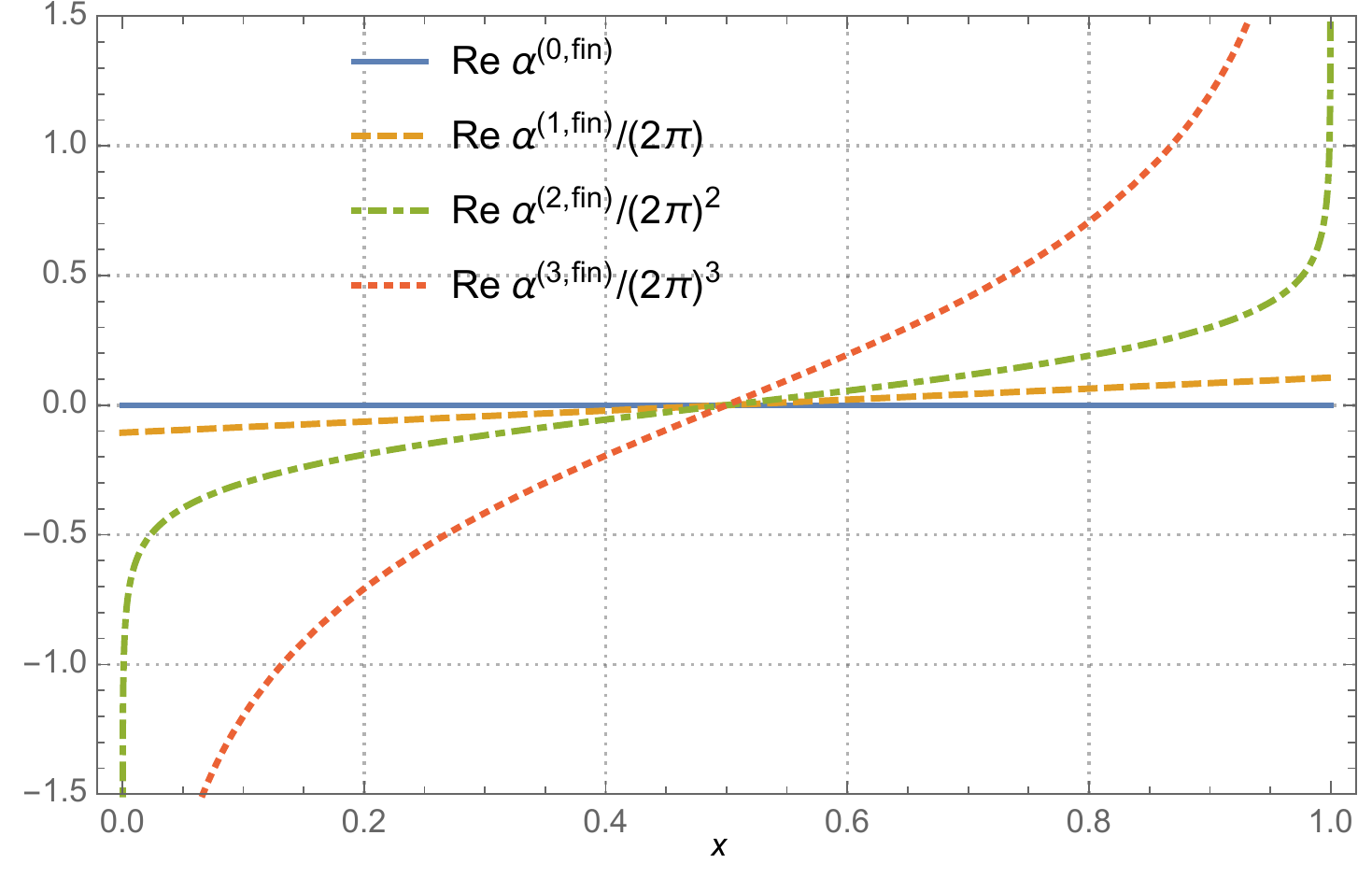}~
    \includegraphics[width=.48\linewidth]{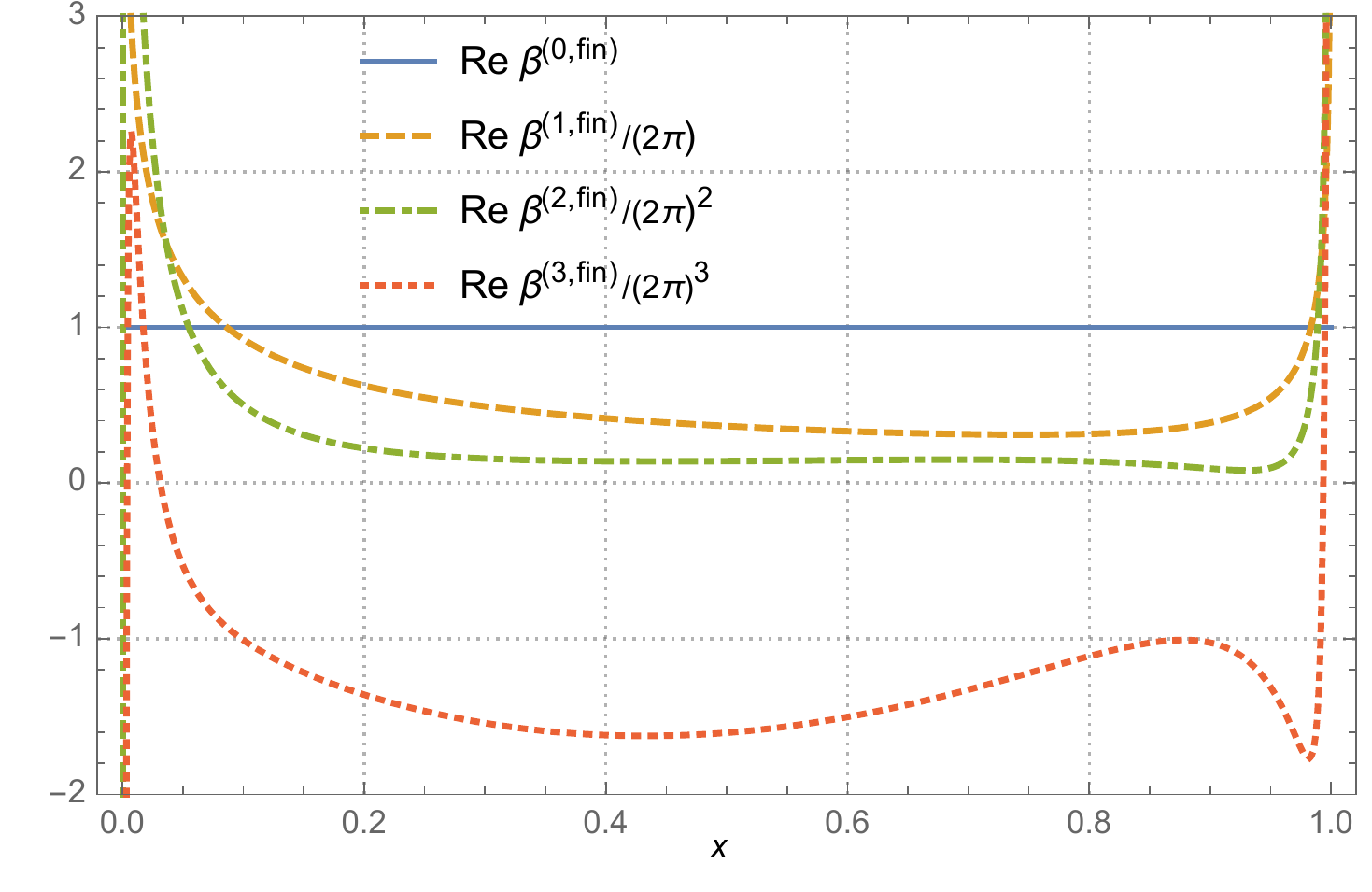}
    \caption{The real part of the three-loop finite remainder functions $\alpha^{(3),\text{fin}}(x)$ and $\beta^{(3),\text{fin}}(x)$ which determine 
    the helicity amplitudes $\mathcal{A}_{L--}$ and $\mathcal{A}_{L-+}$, respectively, for $u\bar{u}\to\gamma\gamma$ and $\mu^2=s$.
    }
    \label{fig:helamp}
\end{figure*}
Using the relations Eqs~(\ref{eq:helamp},\ref{eq:alphabeta}), we obtain the three-loop renormalised finite remainders 
$\alpha^{(3,{\rm fin})}$, $\beta^{(3,{\rm fin})}$, $\gamma^{(3,{\rm fin})}$, 
$\delta^{(3,{\rm fin})}$, defined in analogy to Eqs~(\ref{eq:renF},\ref{eq:catani}). 
This represents the
first three-loop calculation of a full QCD amplitude. 
Our result can be expressed in terms of harmonic polylogarithms of weights $0$ and $1$, or, alternatively, in terms of a more compact functional basis consisting of 14 classical polylogarithms and the 9 functions
\begin{gather}
\Li_{3,2}(x,1),~\Li_{3,2}(1-x,1),~\Li_{3,2}(1,x), \notag \\
\Li_{3,3}(x,1),~\Li_{3,3}(1-x,1),~\Li_{3,3}(x/(x-1),1), \notag \\
\Li_{4,2}(x,1),~\Li_{4,2}(1-x,1),~\Li_{2,2,2}(x,1,1),
\end{gather}
which are two and three-fold nested sums in the conventions of \cite{Vollinga:2004sn}.\footnote{Note that  \texttt{PolyLogTools}~\cite{Duhr:2019tlz} uses 
a different convention for indices and arguments compared to the ones we adopt here.}
In the latter representation, our three-loop amplitudes can be evaluated in fractions of a second in one phase space point.
Although the result is  relatively compact,
it is too long to be presented here. We attach it in electronic format to the 
arXiv submission of this letter.
In Fig.~\ref{fig:helamp} we show a numerical evaluation of our analytic result for all independent finite remainder functions through to three loops.
While only the real parts are shown here, we note that the imaginary parts of the loop corrections can actually exceed their real parts substantially in the case of $\beta$.

Before concluding, we list some of the checks that we have performed to verify
the correctness of our calculation. We have performed
the tree-level, one- and two-loop computation from scratch using our setup, and
obtained agreement for the finite reminders at $\mathcal O(\ep^0)$ with results
in the literature~\cite{Glover:2003cm}. 
From Eq.~\eqref{eq:catani} it  follows, that to extract 
$\oF_{i}^{(3,{\rm fin})}$,
one also requires the one and two loop results expanded up to order $\ep^4$ and $\ep^2$, respectively. 
To check their correctness, we have abelianised the result of~\cite{Ahmed:2019qtg} and checked against
the abelian part of our results up to weight six.
As we have mentioned, Bose symmetry relates
amplitudes of different helicities. 
We have verified that our helicity amplitude coefficients up to three
loops display the Bose symmetry properties in Eq.~(\ref{eq:bosesymm}).

A non-trivial aspect of our calculation is the analytic
continuation of all the required integrals to the physical region. To check our
procedure, we verified that our  solutions are consistent with the boundary values for the master integrals that 
can be inferred by imposing regularity conditions on their differential equations. This allowed us to relate complicated
four-point  integrals to simpler two- and three-point ones (see~\cite{vonManteuffel:2015gxa} for a compilation), whose analytic continuation is straightforward.
Moreover, we used Reduze 2 to 
find non-trivial symmetry relations among the master integrals 
and their crossings and verified that
they are all satisfied by our analytic results.
In addition, we checked the finite part of the two-loop integrals against the literature and 
some of the simple three-loop integrals against secdec~\cite{Borowka:2015mxa} .
Finally, the most powerful check of the correctness of our result is that the remainders 
$\oF_i^{(3,{\rm fin})}$ are in fact finite in the $\ep\to 0$ limit. This also tests in a non-trivial way our
analytic continuation procedure, as it links three-loop integrals with lower loop ones.

In conclusion, in this letter we have reported the first calculation of a three-loop four-point scattering amplitude in full QCD.
We have obtained our results by defining a minimal set of projectors that allowed us to extract all the information
required to reconstruct the helicity amplitudes. Our result can be expressed in terms of classical polylogarithms plus
a handful of multiple polylogarithms, and it is very compact for a QCD amplitude of this type. 
The methods we employed for this calculation
are generic and can be used to compute the three-loop helicity amplitudes for 
$2\to2$ scattering of massless
partons in QCD. We leave this for future investigations.

{\bf Acknowledgments}
We are grateful to P. Bargiela for providing several non-trivial checks 
of the one- and two-loop helicity amplitudes
at higher orders in the dimensional regulator $\ep$. 
LT is grateful to T. Peraro for 
many clarifying discussions on the general idea behind the 
tensor decomposition for scattering amplitudes used in this paper. 
We also thank T. Peraro for various comments 
on the manuscript and B. Mistlberger for clarifications on the results of~\cite{Ahmed:2019qtg,Henn:2020lye}. 
The research of FC is partially supported by
the ERC Starting Grant 804394 HipQCD.
AvM is supported in part by the National Science Foundation through Grants 1719863
and 2013859.
LT is supported by the Royal Society through Grant URF/R1/191125.

\bibliographystyle{bibliostyle}
\bibliography{Biblio}

 \newpage
\onecolumngrid
\newpage
\appendix

\section*{Supplemental material}

\makeatletter
\renewcommand\@biblabel[1]{[#1S]}
\makeatother

\subsection{The projectors}
The projectors defined in Eq.~\eqref{eq:projT} can be found in the following way.
Contracting them with the scattering amplitude in Eq.~\eqref{eq:AT}
and summing over the helicities of the external states in $d$ space-time dimensions, we require
\begin{align}
\sum_{pol} \mathcal{P}_i \mathcal{A}(s,t)= \oF_i(s,t)\,.
\end{align}
Note that following the choice in Eq.~\eqref{eq:gauge}, 
the $d$ dimensional polarisation sum for the photons reads
\begin{equation}
\begin{split}
\sum_{pol} \epsilon_3^{\mu} \epsilon_3^{* \nu} = - g^{\mu \nu} + \frac{ p_2^\mu p_3^\nu + p_3^\mu p_2^\nu }{p_2 \cdot p_3}\,\, \\
\sum_{pol} \epsilon_4^{\mu} \epsilon_4^{* \nu} = - g^{\mu \nu} + \frac{ p_1^\mu p_4^\nu + p_4^\mu p_1^\nu }{p_1 \cdot p_4} \,.
\end{split}
\end{equation}
If we define the matrix
$M_{ij} = \sum_{pol} \oT_i^\dagger \oT_j\,,$
the coefficients of the projectors can be expressed rather compactly as
$c_{ik} = \left(M^{-1}\right)_{ik}$
with
\begin{equation}
M^{-1}= \frac{1}{(d-3)(s+u)} \left(
\begin{array}{cc}
X & 0  \\
0 &  -\frac{1}{2u(d-4)}
\end{array}
\right)\,, \qquad 
X =  \left(
\begin{array}{cccc}
-\frac{u}{2 s^2} & 0 & -\frac{u}{2 s^2 (s+u)} & 0 \\
0 & -\frac{u}{2 s^2} & \frac{u}{2 s^2 (s+u)} & 0 \\
-\frac{u}{2 s^2 (s+u)} & \frac{u}{2 s^2 (s+u)} & -\frac{d u^2+4 s^2+4 s u}{2 s^2 u (s+u)^2} & \frac{2 s+u}{2 s u (s+u)} \\
0 & 0 & \frac{2 s+u}{2 s u (s+u)} & -\frac{1}{2 u} \\
\end{array}
\right)\,.\label{eq:matrixprojqqbaa2}
\end{equation}
As one can observe from Eq.~\eqref{eq:matrixprojqqbaa2}, the fifth tensor $\oT_5$ decouples from the other four.
This, combined with the fact that it is always zero
for four dimensional external states,
is enough to prove that the helicity amplitudes in four dimensions 
receive contribution only from the first four form factors in Eq.~\eqref{eq:AT}, see Eqs~\eqref{eq:alphabeta}.

It is possible to use our projectors to compute the generic  $n$-loop $\times$ $m$-loop contraction in Conventional Dimensional Regularisation (CDR), summed over colours and polarisations. A simple exercise gives
\begin{align}
 \frac{1}{N_c} \sum_{pol,col}\mathcal{A}^{(n)} \mathcal{A}^{(m)*} &=  
  \frac{2 (s-t)t}{u}\, \oF_4^{(n)} \left[
  - s  \oF_1^{(m)*} 
  + s  \oF_2^{(m)*}
  - s t \oF_3^{(m)*}
   -\frac{2 \oF_4^{(m)*} \left(-s^2-t^2+u^2 \epsilon \right)}{(s-t)}
   \right]
   \nonumber \\
   &
   + \frac{2 s t }{u}\, \oF_1^{(n)} \left[
   -2 s  (\epsilon -1) \oF_1^{(m)*}
   - s  \oF_2^{(m)*}
   + s t \oF_3^{(m)*}
   -\oF_4^{(m)*} (s-t)
   \right]
   \nonumber \\
   &
   + \frac{2 s t}{u}\, \oF_2^{(n)} \left[
   - s\oF_1^{(m)*}
   -2 s (\epsilon -1)\oF_2^{(m)*}
   - s t \oF_3^{(m)*}
   + \oF_4^{(m)*} (s-t)\right]
   \nonumber \\
   &
   + \frac{2 s t^2}{u}\, \oF_3^{(n)}
   \left[ s \oF_1^{(m)*}
   - s  \oF_2^{(m)*}
   + s t \oF_3^{(m)*}
   -\oF_4^{(m)*} (s-t)\right]
   \nonumber \\
   &+4 t u\, \epsilon\,  (2 \epsilon -1) \oF_5^{(m)*} \oF_5^{(n)}\,,
\end{align}
where $N_c=3$ is the number of colours.
By direct inspection of this formula it is then obvious that, with our choice of tensors, 
the fifth form factor is only relevant to obtain results
for the higher orders in $\epsilon$ of the scattering amplitude in the CDR
scheme.

\subsection{The infrared structure of the loop amplitude}
In this appendix, we define the explicit formulas for the infrared subtraction operators 
$\mathcal I_i$ required to define our finite
remainders Eq.~\eqref{eq:catani}. We follow the notation of Ref.~\cite{ Becher:2009qa} and
refer the reader to that reference for a detailed discussion of the infrared
structure of loop amplitudes. For our case, we define the following operators
\begin{equation}
  \begin{split}
    &
    \mathcal I_1 = \frac{\Gamma'_0}{4\epsilon^2} + \frac{\Gamma_0}{2\epsilon},
    \\
    &
    \mathcal I_2 = -\frac{\mathcal I_1^2}{2}
    -\frac{\beta_0}{2\epsilon}\left( \mathcal I_1 + \frac{\Gamma'_0}{8\epsilon^2}\right)
    +\frac{\Gamma'_1}{16\epsilon^2}
    +\frac{\Gamma_1}{4\epsilon},
    \\
    &
    \mathcal I_3 = -\frac{\mathcal I_1^3}{3}-\mathcal I_1 \mathcal I_2
    +\frac{\beta_0^2\Gamma'_0}{36\epsilon^4}
    -\frac{\beta_0}{3\epsilon}\left(\mathcal I_1^2+2\mathcal I_2
    +\frac{\Gamma'_1}{12\epsilon^2}\right)
    -\frac{\beta_1}{3\epsilon}\left(\mathcal I_1 +
    \frac{\Gamma'_0}{12\epsilon^2}\right)
    +\frac{\Gamma'_2}{36\epsilon^2}
    +\frac{\Gamma_2}{6\epsilon},
  \end{split}
  \label{eq:poles}
\end{equation}
with
\begin{equation}
  \Gamma'_i = -2 C_F\; \gamma_{c,i}\,,\qquad 
  \Gamma_i =  2\gamma_{q,i}+
  C_F \;\gamma_{c,i}\; \ln\left(\frac{-s_{12}-i0}{\mu^2}\right).
\end{equation}
In Eq.~\eqref{eq:poles} we used
\begin{equation}
\beta_0 = \frac{11}{6}C_A - \frac{2}{3}T_R n_f\,,\qquad 
\beta_{1} = \frac{17}{6}C_A^2 - \frac{5}{3}C_A T_R n_f - C_F T_R n_f\;.
\end{equation}
The perturbative expansion of the cusp anomalous dimension reads
\begin{equation}
  \begin{split}
    \gamma_{c,0} &= 2\,,
    \\
    \gamma_{c,1} &= \left(\frac{67}{9}-\frac{\pi ^2}{3}\right) C_A
    -\frac{20 n_f T_R}{9}\,,
    \\
    \gamma_{c,2} &= C_A^2 \left(\frac{11 \zeta_3}{3}+\frac{245}{12}
    -\frac{67 \pi ^2}{27}+\frac{11
      \pi ^4}{90}\right)+C_A n_f T_R \left(-\frac{28 \zeta_3}{3}
    -\frac{209}{27}+\frac{20 \pi ^2}{27}\right)
    \\
    &
    +C_F n_f T_R
    \left(8 \zeta_3-\frac{55}{6}\right)-\frac{8 n_f^2 T_R^2}{27}\,;
  \end{split}
\end{equation}
while the one for the quark anomalous dimension reads
\begin{equation}
  \begin{split}
    \gamma_{q,0} &= -\frac{3 C_F}{2}\,,
    \\
    \gamma_{q,1} &= C_A C_F
    \left(\frac{13 \zeta_3}{2} -\frac{961}{216}-\frac{11 \pi
      ^2}{24}\right)+C_F^2 \left(-6 \zeta_3-\frac{3}{8}+\frac{\pi
      ^2}{2}\right)+\left(\frac{65}{54}+\frac{\pi ^2}{6}\right) C_F
    n_f T_R\,,
    \\
    \gamma_{q,2} &=
    C_F^3 \left(-\frac{17 \zeta_3}{2}+\frac{2
      \pi ^2 \zeta_3}{3}+30 \zeta_5-\frac{29}{16}-\frac{3 \pi
      ^2}{8}-\frac{\pi ^4}{5}\right)
    \\
    &
    +C_A C_F^2 \left(-\frac{211
      \zeta_3}{6}-\frac{\pi ^2 \zeta_3}{3}-15
    \zeta_5-\frac{151}{32}+\frac{205 \pi ^2}{72}+\frac{247 \pi
      ^4}{1080}\right)
    \\
    &+
    C_A^2 C_F \left(\frac{1763
      \zeta_3}{36} -\frac{11 \pi ^2 \zeta_3}{18}-17
    \zeta_5-\frac{139345}{23328}-\frac{7163 \pi ^2}{3888}-\frac{83 \pi
      ^4}{720}\right)
    \\
    &
    +C_F^2 n_f T_R \left(\frac{64
      \zeta_3}{9}+\frac{2953}{216}-\frac{13 \pi ^2}{36}-\frac{7 \pi
      ^4}{54}\right)+C_F n_f^2 T_R^2 \left(-\frac{4
      \zeta_3}{27}+\frac{2417}{1458}-\frac{5 \pi ^2}{27}\right)
    \\
    &
    +C_A C_F n_f T_R \left(-\frac{241 \zeta_3}{27}
    -\frac{8659}{2916}+\frac{1297 \pi ^2}{972}+\frac{11 \pi
      ^4}{180}\right)\,.
  \end{split}
\end{equation}


\end{document}